# DFT aided prediction of phase stability, optoelectronic and thermoelectric properties of $A_2AuScX_6$ (A= Cs, Rb; X= Cl, Br, I) double perovskites for energy harvesting technology

S. Mahmud[1,2], M.A. Ali[1], M. M. Hossain[1], M. M. Uddin[1]

[1]Advanced Computational Materials Research Laboratory, Department of Physics, Chittagong University of Engineering and Technology (CUET), Chattogram-4349, Bangladesh.

[2]Department of Electrical and Electronic Engineering, Jatiya Kabi Kazi Nazul Islam University (JKKNIU), Mymensingh-2224, Bangladesh.

**Abstract**

In this work, density functional theory (DFT) is used to find out the ground state structures of $A_2AuScX_6$ (A= Cs, Rb; X= Cl, Br, I) double Perovskite (DP) halides for the first time. The DP $A_2AuScX_6$ halides were studied for their structural phase stability and optoelectronic properties in order to identify potential materials for energy harvesting systems. The stability criteria were verified by computing the formation energy, binding energy, phonon dispersion curve, stiffness constants, tolerance, and octahedral factors. The electronic band structure, carrier effective mass, density of states (DOS), and charge density distribution were calculated to reveal the nature of electronic conductivity and chemical bonding nature present within them. For $Cs_2AuScX_6$ ($Rb_2AuScX_6$) [X = Cl, Br, I], the corresponding values of band gap [using TB-mBJ] are 1.88 (1.93), 1.68 (1.71), and 1.30 (1.32) eV. The optical constants (dielectric function, absorption coefficient, refractive index, energy loss function, photoconductivity, and reflectivity) were also calculated to get more insights into their electronic nature. For $Cs_2AuScX_6$ ($Rb_2AuScX_6$) [X = Cl, Br, I], the absorption coefficient in the visible range are 3.33 (3.45) $\times10^5$ cm$^{-1}$, 2.70 (2.81) $\times10^5$ cm$^{-1}$, and 2.13 (2.18) $\times10^5$ cm$^{-1}$, respectively. We also investigated the thermoelectric properties to predict promising applications in thermoelectric devices. Our calculations revealed high *ZT* values of 0.92, 1.07, and 1.06 for $Cs_2AuScX_6$ (X = Cl, Br, I) and 0.97, 0.99, and 1.01 for $Rb_2AuScX_6$ (X = Cl, Br, I) at 300 K. To further aid in predicting any novel materials, routine research was also done on the thermo-mechanical characteristics. The results suggest the compounds considered as potential candidates for use in solar cells and/or thermoelectric devices.

Keywords: Double Perovskite halides; DFT computation; Phase stability; Optoelectronic properties; Thermo-electric properties; Thermo-mechanical properties.

## 1. Introduction

In the last few years, the energy demand has grown quickly, which has made experts work harder to find alternative sources of energy other than the use of fossil fuels along with additional usual sources. Sunlight, which is natural and provides a lot of energy, is a significant power system option[1]. Researchers have paid tremendous attention to developing affordable and dependable materials that can significantly use solar radiation to produce energy. The photovoltaic process represents a viable means of generating renewable energy[2], and additional study is required to predict solar cell materials with superior efficiency.

The rapid increase in energy consumption over the past several years has pushed scientists to identify more conventional and alternative energy sources that do not need the burning of fossil fuels. Naturally occurring and abundant in energy, sunlight is a major power system option[1]. Scientists have worked hard to predict reliable, inexpensive materials that can harness a considerable amount of solar radiation to generate electricity. The photovoltaic method is a feasible way to produce renewable energy[2], but further research is needed to identify solar cell materials with higher efficiency.

A semiconductor called Perovskite is employed to carry the electric charge whenever light strikes the substance. The chemical formula for the Perovskite crystal structure is $ABX_3$, where A and B are cations, and X is an anion. A type of substance known as "double Perovskite" has a Perovskite crystal structure, but two distinct cations occupy the B-site of the structure. A double Perovskite has the formula $A_2BB'X_6$ because the B-site contains two distinct cations. The B and B' cations may alter, but the A-site cation stays the same as in a typical Perovskite. This configuration offers various features and capabilities and permits different cation combinations. The double Perovskite (DP) structure of $A_2AuScX_6$ (A= $Cs^+$/$Rb^+$; X= $Cl^-$/$Br^-$/$I^-$) consists of two interpenetrating face-centered cubic lattices, one formed by $Cs^+$/$Rb^+$ cations and the other by $Au^{1+}$, $Sc^{3+}$, and $X^-$ anions. The choice of Au and Sc as the transition metal cations in each substance are particularly significant. Au is a noble metal known for its excellent electrical conductivity, while Sc is recognized for its tunable bandgap and favorable optoelectronic properties. Combining these two elements in the B-site of the double Perovskite halide makes it possible to tailor the material's electronic band structure and charge transport properties. This tenability is essential for optimizing the material's performance in electronic and optoelectronic devices. The halide component, X (chloride, bromide, or iodide), plays a crucial role in determining the overall properties, especially influencing the electronic band structure and affecting the material's absorption characteristics. Also, by controlling the halide composition, researchers can modulate the energy levels and light absorption properties of materials, further enhancing their potential for optoelectronic applications. The main thing to remember is that DP can support B-site elements with oxidation states ranging from $1^+$ to $4^+$, whereas $ABX_3$ can only handle $2^+$ B-site cations[3]. Moreover, the zero oxidation state, a charge balance typical of Perovskite halides, may be recognized by combining tetravalent ions with a vacancy at BB'. Moreover, the band gap would be direct or indirect with larger values is impractical for solar cell use.

The organic-inorganic mixed or hybrid halide Perovskite $CH_3NH_3PbX_3$ ($MAPbX_3$; X= Cl/Br) has received a lot of interest as a key component in the fabrication of extremely effective solar cells due to its excellent efficiency[4–9]. Scientists have put in a lot of work to increase efficiency from 3.8% in 2009 to 26.08% in 2023[10–14]. The strong absorbance of light in the visible area and outstanding stability of organic-inorganic mixed or hybrid halide perovskite lead to their superior performance[15,16]. However, lead (Pb), which cannot be commercialized on a big scale due to its poor stability and toxicity, necessitates the search for other materials as alternatives for B-sites.

Recently, there has been a lot of excitement in double perovskite (DP) halides due to their potential applications in various fields, including opto-electronics[17–19], magnetism[20–22], catalysis[23–25], and energy conversion[5,26–28]. They also have a range of exciting properties that make them potential for next-generation electronic devices and energy technologies, including magneto-resistance[29], multi-ferroicity[30], and high spin-orbit coupling[31]. The DP halide group creates a new photovoltaic and optoelectronic device, but it is challenging to determine which ones have the optimum electrical, structural, and energetic properties for solar cell applications. Recently, DP has gained importance in optoelectronic studies[32,33] because of its capacity to accept a variety of metal cations. Scientists are now studying DP to develop new materials with enhanced functionality and better understand the structure-property correlations associated with these materials. These materials might eventually lead to the development of more efficient and eco-friendly devices, and they offer a great deal of promise to enhance technology across a range of sectors. For example, in the case, Slavney et al.[32] prepared and examined $Cs_2AgBiBr_6$ double Perovskite and stated that inorganic compositions are interesting choices for optical and solar applications and should be explored extensively. Mcclure[34] experimentally synthesized the $Cs_2AgBiX_6$ (X= Cl/Br) solid solution and found the indirect band gap of 2.77 and 2.19 eV that are somewhat less than the band gaps of the similar lead halide Perovskites, 3.00 eV for $CH_3NH_3PbCl_3$ and 2.26 eV for $CH_3NH_3PbBr_3$. F. Igbari et al.[35] effectively prepared $Cs_2AgBiBr_6$ with an energy gap measured at 1.95 eV, suggesting as a potential photovoltaic material. A DP of $Cs_2AgInCl_6$ with a high photosensitive characteristics and a direct band nature was created by Volonakis et al[36]. In addition, many researchers have used the DFT approach for double halide Perovskite in solar cell, since it is a cost-effective and time-efficient alternative. Dhar et al. studied the lead free DP halide of $Cs_2ScInX_6$ (X= Cl/Br) compound theoretically and found the direct band gap of 0.637 and 0.81ev for optoelectronic device[3,37]. N.E. et al. showed the optoelectronic and thermal properties of $Cs_2ScTlX_6$ (X= Cl/Br/I) for renewable energy applications with the band gap of 3.85 eV (direct), 3.2 (direct), and 2.75 eV (indirect). But these compounds exhibited better thermoelectric properties (figure of merit 0.61 to 0.66) compared to Cs based halide[38]. Recently, Ashiq et al. developed B-site cations of Sc and Au based $K_2ScAuZ_6$ (Z= Br/I) DP halides for harvesting energy appliances with the indirect band gap values of 2.00 and 1.45 eVwith ZT values of 0.80 and 1.00 at 200K[39].

We selected a group of double halide compounds of $A_2AuScX_6$ (A= Cs, Rb; X= Cl, Br, I). Six halides are present here: the alkaline metal Cs/Rb (group 1) and the rare earth soft metal Sc (group 3) are both paramagnetic, while the metal Au (group 11) is diamagnetic. To our knowledge, the aforementioned materials have not been the subject of any theoretical or applied research. Prediction of some new materials with better optoelectronic properties suitable for energy technologies is the primary goal of the present research.

Thus, the stability, optoelectronic characteristics, and thermoelectric properties of $A_2AuScX_6$ halides (where A = Cs, Rb, and X = Cl, Br, I) have been studied in this work. We anticipate that our study will offer vital guidance for predicting lead-free DP materials that may be used in future thermoelectric devices and/or solar cells. In addition, as a routine check, this article also discussed the thermo-mechanical characteristics of the named solids.

## 2. Computational technique

This task was carried out by the Wien2k code[40] using the DFT on the Kohn shams equation. For this, firstly, ground state variables are computed using the Generalized Gradient Approximation (GGA) Perdew-Burke-Ernzerhof (PBE)[41] approach with FP-LAPW (Full potential linearized augmented plane wave)[42] method for structural optimization by establishing the Birch-Murnaghan 3rd state of equation. Initialization was performed using the following parameters: bond length factor of 3, the energy required for valence and core state separation of -6.0 Ry, R-MT*K-MAX of 7, number of employed k-points of 1000 (10×10×10), and GMAX of 12. The DFT calculation in Wien2k involves SCF (self-consistent field), which solves the Kohn Shams equation by updating the charge density on the electronic wave function. When calculating SCF, the energy convergence (ec) and charge convergence (cc) criteria are set at 0.0001Ry and 0.001e, respectively. Thermal and mechanical investigations are also conducted using the IR-Elast program[43]. In order to deliver more accurate results, we used the Trans Blaha modified Becke-Johnson (TB-mBJ)[44] methodology for optoelectronic characteristics since it maintains a balance between precision and computing time for inorganic compounds. The Boltztrap2 algorithm[45] was used to determine the behavior of thermal electronic transport, with the exception of lattice vibration. The 1×1×1 supercell concept was used to perform the calculation of phonons using the technique of finite displacement way. For this computation, we used the following electron-valence configurations: $5p^6 6s^1$ for the Cs, $4p^6 5s^1$ for the Rb, $5d^{10} 6s^1$ for the Au, $4s^2 3d^1$ for the Sc, $3s^2 3p^5$ for the Cl, $4s^2 4p^5$ for the Br, and $5s^2 5p^5$ for the I.

## 3. Results and discussion

3.1 Structural properties and stability

Figure 1 depicts the cubic crystal structure of $A_2AuScX_6$ (A= Cs, Rb; X = Cl, Br, I) double perovskite (DP) halides, where the four formula units are present in the ratio of 2:1:1:6, and are comparable to the basic perovskite of $ABX_3$. The DP structures are made up of 8 Cs/$R_b$ atoms, 13 [Au-X6] octahedra and 14 [Sc-X6] octahedra. These are all members of the face-centered Fm-3m (225) symmorphic space group and are located in the Wyckoff positions such as 8c, 4a, 4b, and 24e of the corresponding atoms Cs/$R_b$ (0.25, 0.25, 0.25), Au (0.5, 0.5, 0.5), Sc (0, 0, 0), and Cl/Br/I (0.25, 0, 0). In this work, the lattice constant increases consistently as the halide atom transitions from Cl to Br to I. The crystal structure was optimized using the GGA-PBE (for

Wien2k) function. Table 1 displays the estimated lattice parameter with the lowest energy at the optimum volume

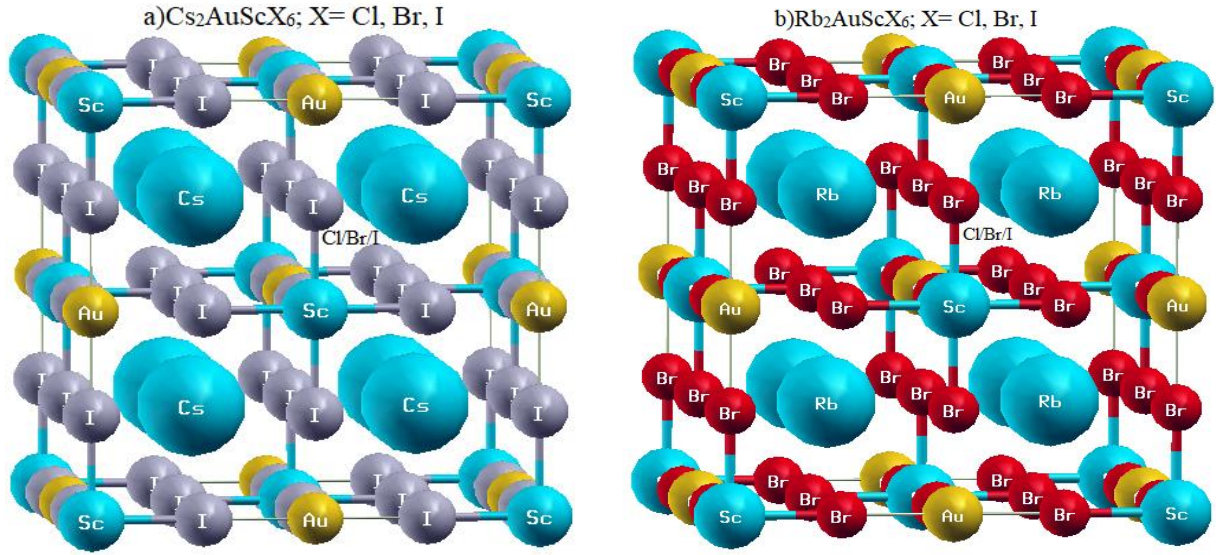

Fig.1: Unit cell of a) $Cs_2AuScX_6$ and b) $Rb_2AuScX_6$; (X= Cl, Br, I)

The DP structural materials for cubic system's stability are determined by tolerance factor ($t_G$) and octahedral factor (μ) which was established by Goldsmith[46] and the following equations are given below:

$$t_G = \frac{R_A + R_X}{\sqrt{2}(\frac{R_{B'}+R_{B''}}{2} + R_X)}$$

$$\mu = \frac{R_{B'} + R_{B''}}{2R_X}$$

Where R represents the efficient ionic radii of each element in the above expressions. So that $R_A$= 1.88 and 1.72 for $Cs^+$ and $Rb^+$, $R_{B'}$ = 1.37 for $Au^+$, $R_{B''}$ = 0.745 for $Sc^{3+}$ and $R_X$ = 1.81, 1.96, 2.2 for $Cl^-$, $Br^-$, and $I^-$ which were taken by the Shannon ionic radius[47]. Table 1 presents the obtained tolerance and octahedral factors of titled materials, and these results witness the stable structures within the range of 0.81 to 1.1 and 0.41 to 0.89[48], respectively.

Furthermore, the formation energy and binding energy serve as further confirmation for the dynamic stability of the DP halides of $A_2AuScX_6$, which is calculated by the following equations[49]:

$$E_f = \frac{E_{A_2AuScX_6} - n_A \times \frac{E_A}{k} - n_{Au} \times \frac{E_{Au}}{l} - n_{Sc} \times \frac{E_{Sc}}{m} - n_X \times \frac{E_X}{p}}{N}$$

$$E_b = E_{A_2AuScX_6} - n_A \times \mu_A - n_{Au} \times \mu_{Au} - n_{Sc} \times \mu_{Sc} - n_X \times \mu_X$$

Where, $E_{A_2AuScX_6}$ denotes the total energy of $A_2AuScX_6$. $E_A$, $E_{Au}$, $E_{Sc}$, $E_X$ and $n_A$, $n_{Au}$, $n_{Sc}$ and $n_X$ denotes the energy and number of atoms of Cs/Rb, Au, Sc, and Cl/Br/I, respectively. The coefficients k, l, m, p are the number of individual atoms per unit cell, and N is the total no of atoms in compounds. The $\mu$ represents the individual Free State energy of the total atom. The negative sign of formation energy confirmed their stability; higher negativity shows greater stability. Consequently, A$_2$AuScCl$_6$ is more stable than A$_2$AuScBr$_6$ and A$_2$AuScI$_6$ for the respective Cs and Rb-based halide.

To evaluate the mentioned compound's thermodynamic or chemical stability, we compute its decomposition energy by considering potential pathways. For this purpose, we use the experimentally identified stable phases of CsAuCl$_3$[50], CsAuBr$_3$[51], CsAuI$_3$[52], CsI[53], RbI[54], ScI$_3$[55], Au[55] and proto-type of Cs$_3$Cr$_2$Cl$_9$[56] for Cl and Br based phase compound. The potential decomposition routes from OQMD for three halides based on Cs and Rb are as outlined below[55]:

$$Cs_2AuScCl_6 = \frac{1}{4} CsAuCl_3 + \frac{7}{10} Cs_3Sc_2Cl_9 + \frac{1}{20} Au$$

$$Cs_2AuScBr_6 = \frac{1}{4} CsAuBr_3 + \frac{7}{10} Cs_3Sc_2Br_9 + \frac{1}{20} Au$$

$$Cs_2AuScI_6 = \frac{1}{4} CsAuI_3 + \frac{3}{10} CsI + \frac{2}{5} ScI_3 + \frac{1}{20} Au$$

$$Rb_2AuScCl_6 = \frac{1}{4} RbAuCl_3 + \frac{7}{10} Rb_3Sc_2Cl_9 + \frac{1}{20} Au$$

$$Rb_2AuScBr_6 = \frac{1}{4} RbAuBr_3 + \frac{7}{10} Rb_3Sc_2Br_9 + \frac{1}{20} Au$$

$$Rb_2AuScI_6 = \frac{1}{4} RbAuI_3 + \frac{3}{10} RbI + \frac{2}{5} ScI_3 + \frac{1}{20} Au$$

The decomposition enthalpies calculated for $Cs_2AuScCl_6$, $Cs_2AuScBr_6$, and $Cs_2AuScI_6$ are 23, 18, and 10 meV/atom, while for $Rb_2AuScCl_6$, $Rb_2AuScBr_6$ and $Rb_2AuScI_6$, they are 19, 15, and 8 meV/atom, respectively. The positive decomposition enthalpy values demonstrate these compound's thermodynamic stability, showing that the energy is derived from the compound's decomposing phases. The third-order Birch-Murnaghan equation of state[57], which supports cubic stability with non-magnetic characteristics and lowest energies, is used in the present samples to optimize the structural or energy volume of the named DPs. A solid's volume and the pressure it experiences are related by the Birch-Murnaghan 3rd-order equation of state, which is given below-

$$E(V) = E_0 + \frac{9V_0 B_0}{16} + \left[\left\{\left(\frac{V_0}{V}\right)^{2/3} - 1\right\}^3 B_0' + \left\{\left(\frac{V_0}{V}\right)^{2/3} - 1\right\}^2 - \left\{6 - 4\left(\frac{V_0}{V}\right)^{2/3}\right\}\right]$$

Where E(V) stands for the internal energy of the material, $E_0$ stands for minimum or ground state energy, $V_0$ stands for the reference or standard volume, V stands for the volume under stress, $B_0$ stands for zero pressure bulk modulus, and $B_0'$ stands for pressure modified bulk modulus. These substances have positive $B_0'$ values, indicating that they become stiffer when subjected to higher pressure. The ground state energy's negative value also represents the stability and growth of that stability for denser materials. The energy vs volume curve of the materials under investigation is shown by supplementary information (see Fig.1S). We have also determined the lattice constant (a) of all compounds from the E-V fitting curve. The atomic radius increases the lattice parameter in compositions based on Cl, Br, and I, and these shifts demonstrate the validity and accuracy of the study.

Table1: Optimized lattice parameter, volume, and structural entity with Fm-3m space group (#225) in non-polarized (NP) states of $A_2AuScX_6$ (A= Cs, Rb; X= Cl, Br, I).

| Parameter | $Cs_2AuScCl_6$ | $Cs_2AuScBr_6$ | $Cs_2AuScI_6$ | $Rb_2AuScCl_6$ | $Rb_2AuScBr_6$ | $Rb_2AuScI_6$ |
|---|---|---|---|---|---|---|
| a=b=c (Å) | 10.5326 | 11.0869 | 11.8920 | 10.4480 | 11.0197 | 11.8452 |
| a=b=c(Bohr) | 19.9037 | 20.9513 | 22.4726 | 19.7438 | 20.8243 | 22.3841 |
| V (Bohr³) | 1989 | 2290 | 2835 | 1924 | 2257 | 2788 |
| $t_G$ | 0.90 | 0.89 | 0.87 | 0.87 | 0.86 | 0.85 |
| μ | 0.58 | 0.53 | 0.48 | 0.58 | 0.53 | 0.48 |
| $E_F$ (eV/atom) | -2.38 | -2.00 | -1.56 | -2.34 | -1.95 | -1.50 |
| $E_b$ (eV/atom) | -4.03 | -3.65 | -3.25 | -4.01 | -3.63 | -3.22 |
| $E_0$ (Ry) | -76324.63 | -102066.17 | -156214.92 | -57089.40 | -82830.94 | -136979.68 |
| $B_0$ | 34.38 | 32.38 | 23.20 | 34.87 | 30.75 | 24.52 |
| $B_0'$ | 6.03 | 5.52 | 4.42 | 5.39 | 4.40 | 3.74 |

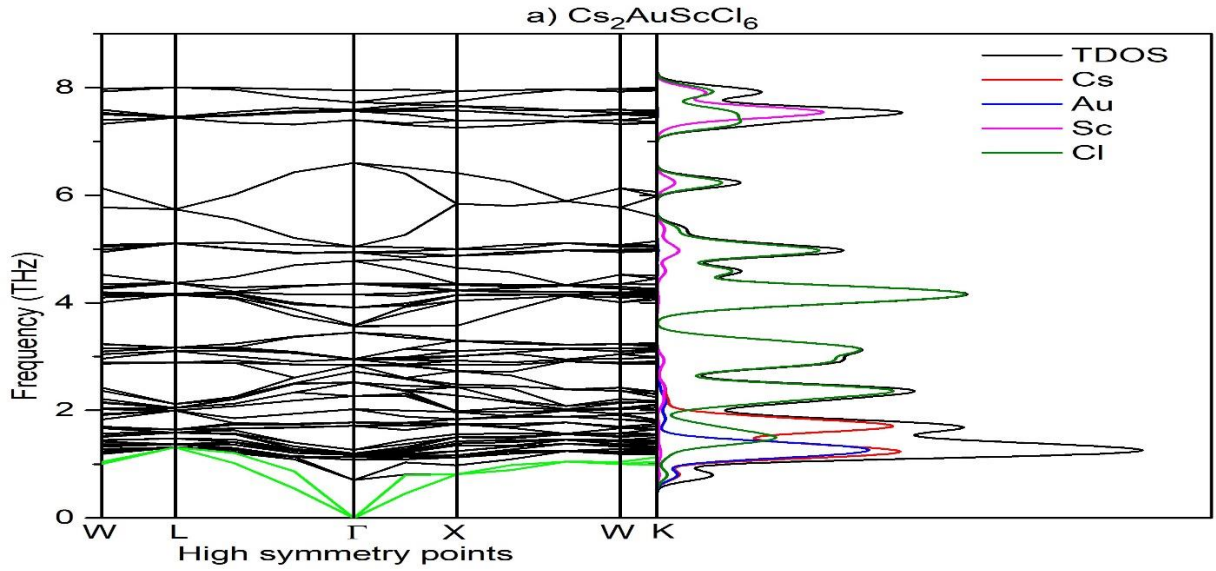
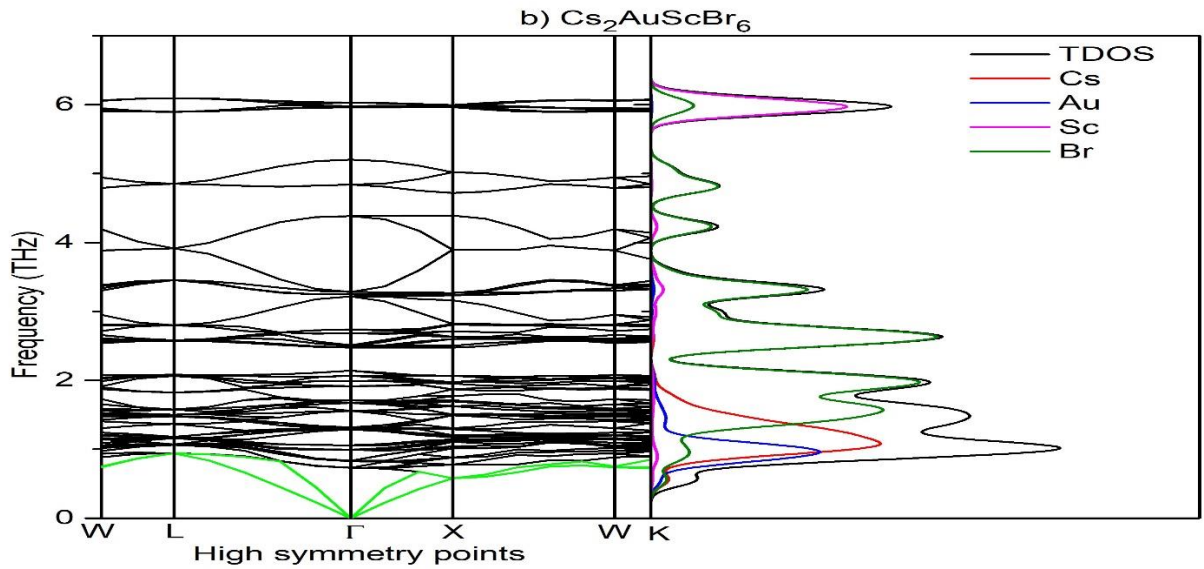
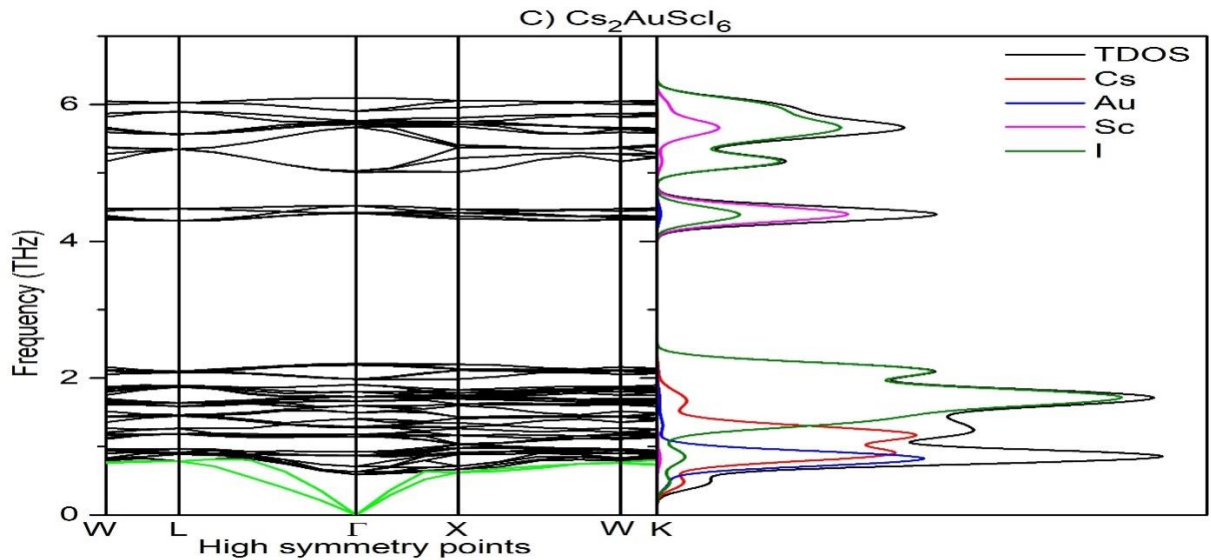

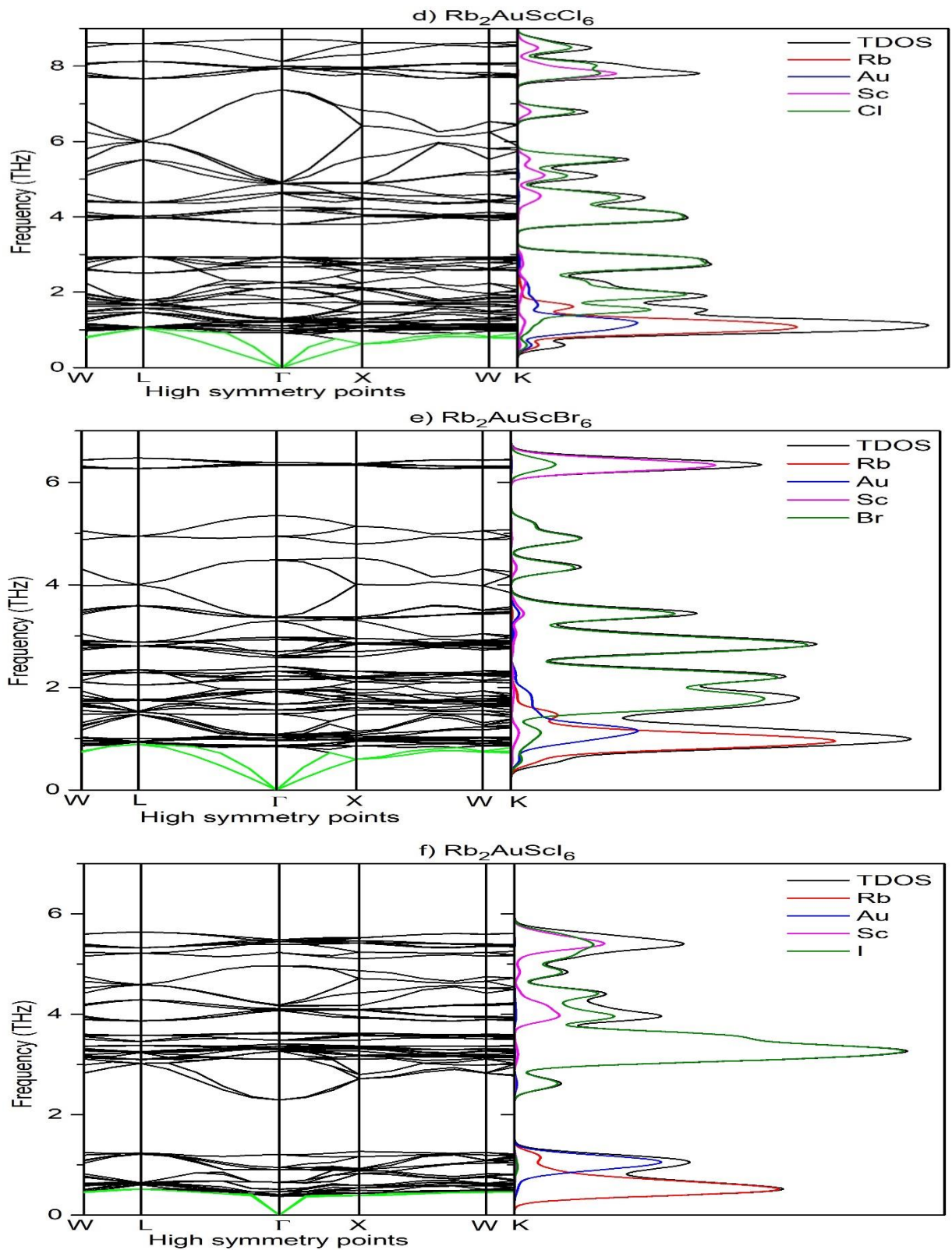

Fig.2: Phonon dispersion with DOS of a) $Cs_2AuScCl_6$, b) $Cs_2AuScBr_6$, c) $Cs_2AuScI_6$, d) $Rb_2AuScCl_6$, e) $Rb_2AuScBr_6$, f) $Rb_2AuScI_6$.

The phonon dispersion curve (PDC) of $A_2AuScX_6$ (A= Cs, Rb; X= Cl, Br, I) with PDOS is depicted in Fig. 2. The PDCs lack of negative frequencies in the band dispersion, where acoustic modes converge at gamma symmetry stations, supports the dynamical equilibrium. The dynamic stability of all drugs under investigation is demonstrated by the lack of an imagined mode during the whole BZ. With 10 atoms, these unit cell compositions must have 3N, or 30 vibrational modes, of which three are acoustic and operate in the low-frequency region (less than 1 THz), and the remaining 27 are optic and operate in the high-frequency range (greater than 1 THz). These are known as zero point frequency and non-zero point frequency, respectively. All compound's phonon particles are stable along the following routes: W to L, L to Γ, Γ to X, X to W, and W to K. The compound's phonon DOS (PDOS) is displayed next to the dispersion curve for better understanding. We observed a strong peak in PDOS for flat bands and a negligible peak for non-flat bands. Both the phonon DOS and the frequency dispersion are precisely adjusted. It is shown that the low-frequency (below 2 THz) lattice vibrations coincide with the optic modes and are mainly ascribed to contributions from Cs and Au atoms.

In contrast, the halide ions (for Cl and Br) dominate the mid-level vibrations, whereas Sc ions work in tandem with Cl/Br ions to dominate the high-frequency area in Cs and Rb-based double Perovskite (DP) halides. However, for $Cs_2AuScI_6$ and $Rb_2AuScI_6$, there is a phononic gap in the middle-frequency range, and I ions work in concert with Sc to generate high-frequency vibrations. As shown in the supplementary part (see Table S1), the poor lattice thermal conductivity of the investigated compounds was caused by the restriction of the phonon dispersion heat resulting from the overlap of acoustic and optical modes in the low-frequency ranges.

**3.2. Electronic properties and the effective mass of charge carrier**

The carrier transport channel may be seen visually through the features of conductors, semiconductors, and insulator's electronic characteristics. Further information about a material's intended applications may be gleaned from its electrical characteristics, such as its band structure and the distribution of electrons inside those bands (known as Total Density of States or Partial Density of States)[58]. The $A_2AuScX_6$ (A= Cs, Rb; X= Cl/Br/I) material's band gap values are essential in deciding whether or not they may be used as photoelectric, optoelectronic, or any other type of particular purpose switching materials. We used the TB-mBJ potential[59] to forecast the band gap with accuracy. Every material has an empty Femi level, which indicates that it possesses a semiconducting nature. In addition, the materials are non-magnetic and show reductions (shifts) in the band structure at the Fermi level when Br or I replace Cl, the halide component.

The GGA-PBE functional technique was first used to investigate the electronic band structure. GGA-PBE + SOC was then employed to see whether there was any effect on the band arrangement. The band structure calculated using these functional methods revealed that the SOC impact on both compounds was very small (less than 0.04). The band structure of all materials under study is displayed in Figs. 3 and 4, utilizing the GGA-PBE (left side) and TB-mBJ (right side) approaches. Here, we present a comparative study of three methods for measuring the bandgap: the well-known TB-mBJ technique, the SCF with SOC, and the traditional SCF.

According to calculations made using the GGA-PBE (TB-mBJ) potential, the energy gaps for $Cs_2AuScCl_6$, $Cs_2AuScBr_6$, $Cs_2AuScI_6$, $Rb_2AuScCl_6$, $Rb_2AuScBr_6$, and $Rb_2AuScI_6$ are 1.65 (1.88), 1.43 (1.68), 0.95 (1.30), 1.69 (1.93), 1.44 (1.71), and 0.97 (1.32) eV, respectively. The fact that the band gap of such substances decreases as the number of atoms of the X element grows lends credence to the idea of the present pattern of oscillations in the energy band gap. The bandgap of the titled materials and a few other similar materials using TB-mBJ are compared in Table 2, which further supports the validity of the computed results for the materials under investigation. Every compound shows the same pattern, and TB-mBJ corrects the GGA-PBE's inaccuracy. The high symmetric points of W (0.50, 0.25, 0.75), L (0.5, 0.5, 0.5), Γ (0, 0, 0), X (0.5, 0, 0), W (0.50, 0.25, 0.75), and K (0.375, 0.375, 0.75) were the destinations of the Brillion zone route. The top of the VB route and the bottom of the CB is more parabolic in shape in the GGA-PBE and TB-mBJ approaches.

The typical energy ranges of perovskite bands are from 0.8 to 2.2eV[60], and materials for photoelectric conversion (PEC) will be able to use this energy extensively. For PEC, the DP of $A_2AuScX_6$ (A= Cs/Rb; X = Cl/Br/I) have suitable band gap values, which have great promise to be used as highly photo-sensitive materials in the not-too-distant future. From Fig. 3 and 4, it is clear that the maximum of valence band (MVB) point and minimum of the conduction band (MCB) point occurs at various symmetry points (L→X for all studied compound by TB-mBJ approach; L→Γ only for $Cs_2AuScI_6$ and $Rb_2AuScI_6$ by GGA-PBE approach), indicating that all DP have indirect band gaps that are semiconducting.

Table 2: Computed band gaps for Cs-based halide of $Cs_2AuScCl_6$, $Cs_2AuScBr_6$, $Cs_2AuScI_6$, and Rb-based halide of $Rb_2AuScCl_6$, $Rb_2AuScBr_6$, and $Rb_2AuScI_6$.

| Compound | GGA-PBE | TB-mBJ | Nature | Reference |
|---|---|---|---|---|
| $Cs_2AuScCl_6$ | 1.65 | 1.88 | Indirect | This |
| $Cs_2AuScBr_6$ | 1.43 | 1.68 | Indirect | This |
| $Cs_2AuScI_6$ | 0.95 | 1.30 | Indirect | This |
| $Rb_2AuScCl_6$ | 1.69 | 1.93 | Indirect | This |
| $Rb_2AuScBr_6$ | 1.44 | 1.71 | Indirect | This |
| $Rb_2AuScI_6$ | 0.97 | 1.32 | Indirect | This |
| $K_2AuScBr_6$ | - | 2.00 | Indirect | [39] |
| $K_2AuScI_6$ | - | 1.45 | Indirect | [39] |

The band gaps estimated from the electronic bannnd structure are less than that obtained from the absorption spectrum for indirect band gap semiconductors[61]. Because the band gap values of $A_2AuScX_6$ (X = Cl/Br/I; A = Cs/Rb) lie within the visible region of the EM (electromagnetic) spectrum, they can be used in solar cell devices. A recent report revealed a significant aspect of the indirect band gap semiconductors for photocatalytic applications owing to the lower radiative recombination rate[62]. Additionally, flat bands inside the CB are observed, indicating the existence of the second phase and potentially explaining the emergence of superconductivity[62]. This suggests that DP is a viable alternative for advanced uses in ultra-powerful magnet technology. The bands still have a detectable range but get broader when Cl replaces the halide Br/I atom.

Solar cells commonly employ silicon (Si) and germanium (Ge) because of their availability, self-passivation, and indirect band gap. Both a photon and a phonon are needed for an indirect band gap semiconductor to go from the valence band (VB) to the conduction band (CB), causing changes in momentum and energy. Since the two variables are inversely related, an ideal characteristic of an efficient solar cell is that the minority carrier diffusion length is higher than the absorber depth of the material[63]. Despite lower light absorption in indirect semiconductors, they often have a longer recombination lifetime and a larger diffusion length, making them viable for efficient solar cells with a sufficiently thick absorbing layer.

So, indirect band gap semiconductors are mostly preferred in thin film solar cells due to their weaker light absorption, enabling a broader photon energy range, including lower-energy solar photons[64–66]. Moreover, materials with an indirect band gap are more defect-tolerant and generate less heat, crucial for solar cell durability. In concentrator photovoltaic, this heat reduction is especially advantageous. Additionally, tandem solar cells combine multiple materials with different band gaps for increased efficiency and can leverage indirect band gap materials to optimize energy conversion across the solar spectrum[32,67,68].

The term "effective mass of charge carriers" describes the mass of electrons or holes in a double Perovskite halide. Their effective mass is essential in determining how charge carriers behave and react to outside stimuli like an electric field. The effective mass in DP can vary depending on the specific material and its composition. It is influenced by the material's band structure, which, in turn, depends on the types of halides and other elements present in the Perovskite structure.

We have calculated the effective mass of charge carriers (electrons or holes) of DP of $A_2AuScX_6$ at high symmetry points of VBM and CBM. The formula used to determine the effective mass and hole by using E-K dispersion of respective materials is as follows:

$$m^* = \frac{\hbar^2}{(d^2E/dK^2)}$$

Where, $\hbar = 1.05 \times 10^{-34}$ J/s. From the E-K dispersion curve, we get the values of $d^2E/dK^2$ by using E-K curve fitting of the respective parabolic curve at any symmetry points. The computed effective mass values are presented in Table 3. It is obvious that the predicted effective mass values are far lower than those reported by others[61] for other DPs. The reduced effective mass is advantageous for transferring carriers, which is very beneficial for solar materials. Therefore, the DPs of $A_2AuScX_6$ are all highly perfect photovoltaic materials and have much promise for use in various solar applications.

Table 3: Estimated effective masses of holes ($m_h^*$) and electrons ($m_e^*$) at the lowest point of CB (L) and Highest point of VB (X or Γ) of respective materials of $A_2AuScX_6$ (A= Cs, Rb; X= Cl, Br, I).

| Compound | Approach | $m_h^*/m_0$ | $m_e^*/m_0$ | $m_h/m_e$ |
| --- | --- | --- | --- | --- |
| $Cs_2AuScCl_6$ | GGA-PBE | 0.32, 0.62[a] | 0.15, 0.47[a] | 2.13 |
|  | TB-mBJ | 0.33, 0.61[a] | 0.18, 0.46[a] | 1.83 |
| $Cs_2AuScBr_6$ | GGA-PBE | 0.28 | 0.14 | 2.00 |
|  | TB-mBJ | 0.28 | 0.15 | 1.86 |
| $Cs_2AuScI_6$ | GGA-PBE | 0.25 | 0.063 | 3.96 |
|  | TB-mBJ | 0.25 | 0.14 | 1.78 |
| $Rb_2AuScCl_6$ | GGA-PBE | 0.31 | 0.15 | 2.06 |
|  | TB-mBJ | 0.32 | 0.18 | 1.77 |
| $Rb_2AuScBr_6$ | GGA-PBE | 0.27 | 0.13 | 2.07 |
|  | TB-mBJ | 0.27 | 0.14 | 1.92 |
| $Rb_2AuScI_6$ | GGA-PBE | 0.25 | 0.064 | 3.90 |
|  | TB-mBJ | 0.25 | 0.13 | 1.92 |

[a]Reference[61] for $Cs_2AgAsCl_6$

Figure 6 shows the titled compound's total and partial DOS and the key band structure characteristics found by the TB-mBJ potential. It offers exact information as well as the involvement of every component. It is clear that every DP compound has the same atomic state characteristics and ways of contribution. 0 eV was chosen as the Fermi level. Due to the little contribution of X-p orbitals, the Sc-d orbitals with hybridization (molecular orbital bonding) account for the majority of the MCB (minimum of conduction band) point of $A_2AuScX_6$. In DOS, the energy value of the band structure and the near $E_F$ of CB match exactly. Au-*d* orbitals contribute the most to the maximum of the valence band (MVB) point, whereas X-*p* orbitals contribute the least. The band structure of PDOS and TDOS ultimately reflects the difference in energy levels between them. Rather than adding much to the band structure, the $Cs^+/Rb^+$ cation close to $E_F$ mostly donates charges to preserve structural stability.

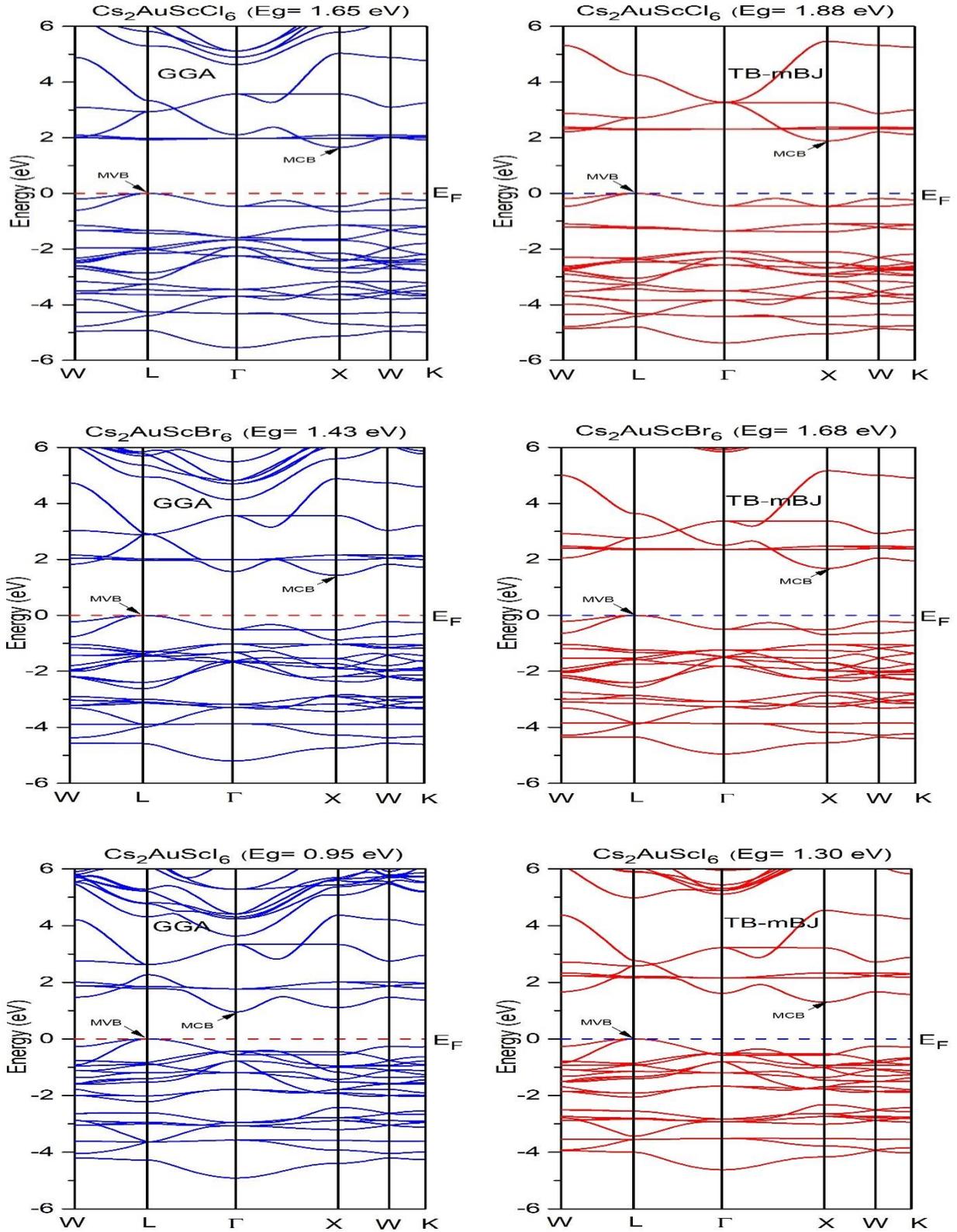

Fig.3: Energy band diagram of Cs-based halide of $Cs_2AuScCl_6$, $Cs_2AuScBr_6$, $Cs_2AuScI_6$ computed by GGA-PBE (Left side) and TB-mBJ (Right side) approach.

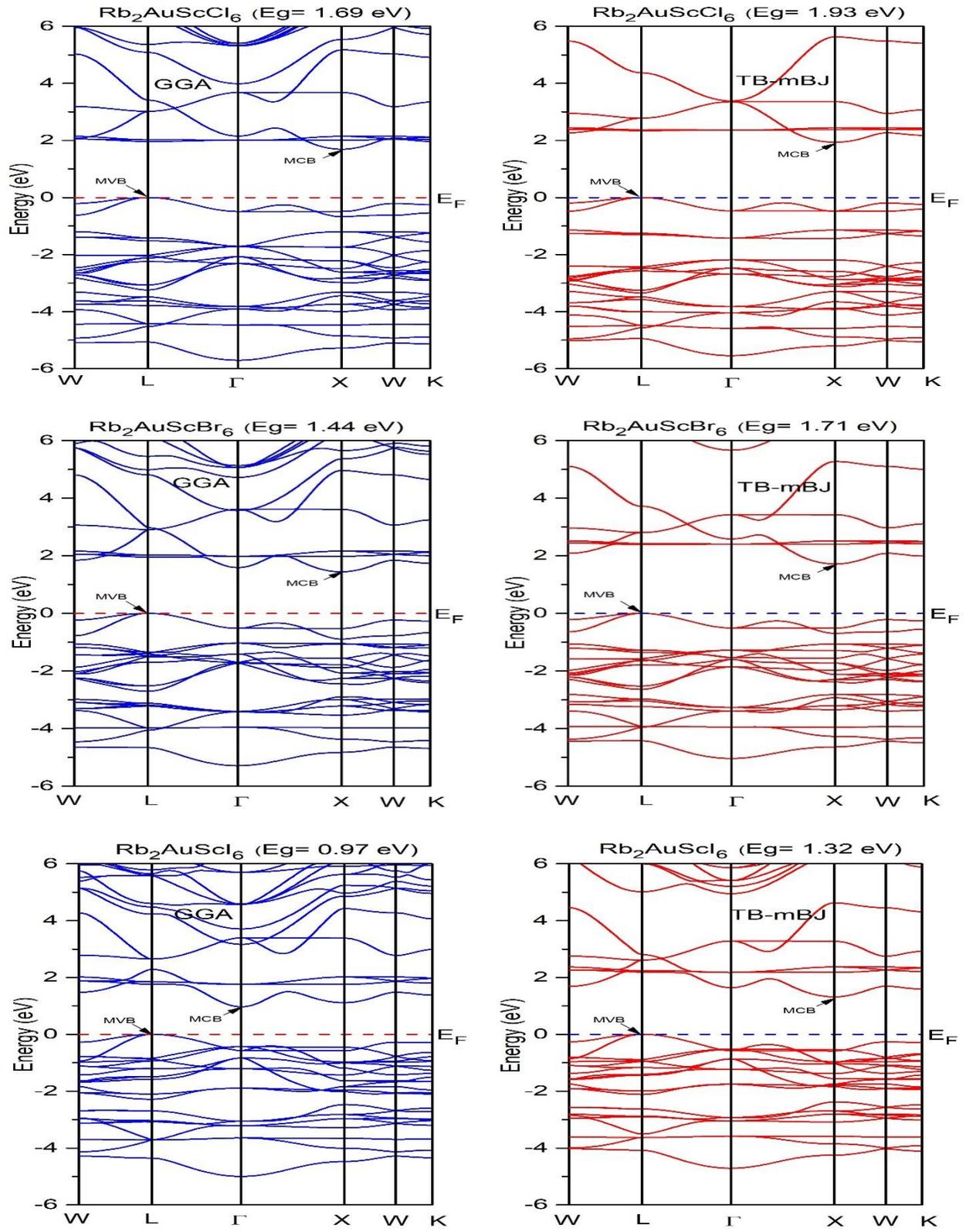

Fig.4: Energy band diagram of Rb-based halide of $Rb_2AuScCl_6$, $Rb_2AuScBr_6$, and $Rb_2AuScI_6$ computed by GGA-PBE (Left side) and TB-mBJ (Right side) method.

## 3.3. Nature of bonding

$A_2AuScX_6$ (A= Cs/Rb; X= Cl/Br/I) compounds belong to the DP family that have intriguing features because of their distinct bonding patterns and structure. It belongs to the class of inorganic DP, and the cations ($Cs^+/Rb^+$) and lattice have an impact on its characteristics. These material's charge density and bonding type can be examined regarding their crystal structure and internal electronic interactions. Space group Fm-3m characterizes the cubic crystal structure of the chemical mentioned. The X ions (Cl, Br, or I) are positioned in the center of these octahedra in this structure, whereas the Au and Sc ions are situated in the corner, sharing octahedral positions. The spaces in between these octahedral $Cs^+/Rb^+$ ions are interstitial.

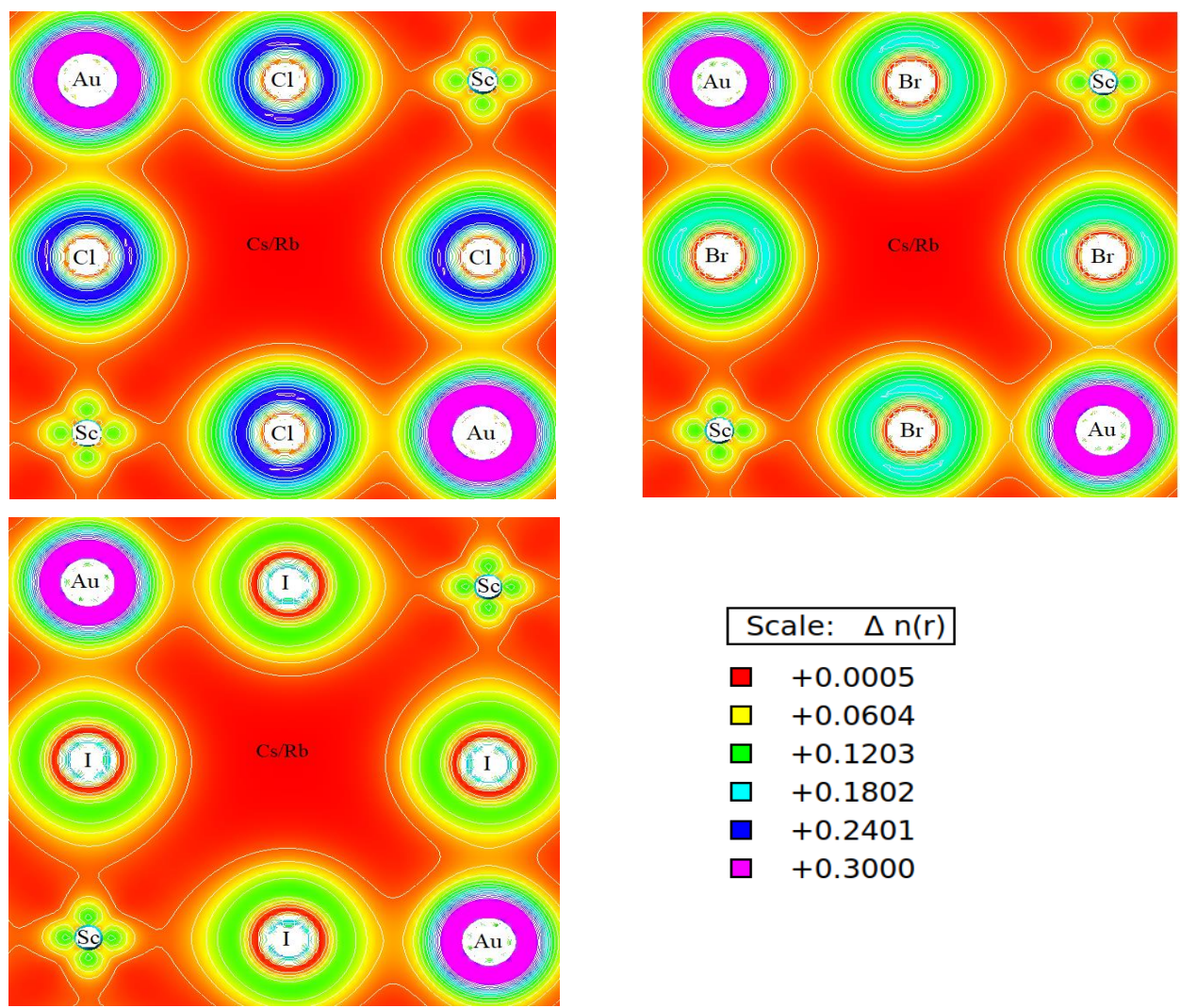

Fig.5: Contour plot of charge density of a) $A_2AuScCl_6$, b) $A_2AuScBr_6$, and c) $A_2AuScI_6$; [A= Cs, Rb].

This configuration creates a network of corner-sharing octahedra, which affects the compound's electronic and bonding characteristics. The charge density contour map for the materials in question is shown in Fig. 5.

The A ions (Cs/Rb) have relatively high electronegativity compared to Au and Sc, leading to ionic bonding between the A ions and the surrounding X ions. This results in forming A-X bonds with a certain degree of ionic character. The Au-Sc-X octahedral involves covalent bonding between the metal ions (Au and Sc) and the halogen ions (X). The electronegativity difference between the metal ions and X ions contributes to polar covalent bonds within the octahedral framework, similar to the reports published[61,69]. The metal-metal bonding (Au-Sc) within the octahedral can also have a metallic character due to the overlap of their atomic orbitals.

The combination of ionic, covalent, and metallic bonding interactions influences the charge density distribution in materials. The A ions contribute to the charge density by providing localized positive charge, while the X ions contribute localized negative charge due to electronegativity. In summary, the charge density and bonding nature of $A_2AuScX_6$ (A= Cs/Rb; X= Cl/Br/I) result from the complex interplay between ionic, covalent, and metallic bonding interactions within its crystal structure. This unusual bonding is the key to many of their desirable photovoltaic properties.

### 3.4. Optical properties

Optical characteristics reflect electronic properties and indicate how a material reacts to incident light. Photon energy in the region of 0 to 14 eV is used to study optical constants such as dielectric function, refraction index, absorption coefficient, and reflectivity and photoelectric/device applications of these DP materials. Their absorption coefficients and the material's electric polarization performance are the primary features of the photovoltaic application.

The complex dielectric function may be represented by the real and imaginary components of the dielectric function with respect to photon energy, which is given by the following formula: $\epsilon(\omega) = \varepsilon_1(\omega) + i\varepsilon_2(\omega)$

The real portion of $\varepsilon_1(\omega)$ indicates the amount of polarization when the external field is applied, can be calculated by the equations of Kramers-Kronig[70], while $\varepsilon_2(\omega)$ indicates an amount of electronic change or light absorption and can be calculated by Kohn-Shams[71] equation, respectively.

$$\varepsilon_1(\omega) = 1 + \frac{2}{\pi} P \int_0^\infty \frac{\omega' * \varepsilon_2(\omega')}{\omega'^2 - \omega^2} d\omega$$

$$\varepsilon_2(\omega) = \frac{Ve^2}{2\pi \hbar m^2 \omega^2} \int d^3k \sum_{nn'} |(kn|p|kn')|^2 f(kn) \times (1 - f(kn')\delta(E_{kn} - E_{kn'} - \hbar\omega)$$

Where P in the first equation stands for the main component, the 2$^{nd}$ equation uses V, e, $\hbar$, p, kn, and kn' to represent unit volume, charge, decreasing plank constant, momentum transfer matrix, VB wave function, and CB wave function.

The dielectric function of the mentioned DP of $A_2AuScX_6$ is shown in Fig.6 (a and b). The real part of the dielectric constant $\varepsilon_1(\omega)$ first experienced a dramatic increase until it reached its maximum value, after which it began a downward trend. The static dielectric constants are 3.27 (3.19), 3.84 (3.76), and 4.97 (4.90) for $A_2AuScX_6$ [A = Cs, Rb; X = Cl, Br, I], and the highest value is seen to be 6.06 (6.03) at 2.33 (2.37) eV, 6.39 (6.36) at 2.22 (2.25) eV, and 7.70 (7.67) at 1.85 (1.89) eV for $A_2AuScX_6$ [A = Cs, Rb; X = Cl, Br, I], respectively. The compounds of Cs-based halide display slightly greater dielectric behavior (static or dynamic) than Rb-based halide. The minimal variation in $\varepsilon_1(\omega)$ for all compounds under study is, however, within -1.16 to -2 for energies greater than 10 eV, where the minus sign of $\varepsilon_1(\omega)$ indicates the compound's metallic character. Also, it was found that switching Cl with Br and I, elevated the strength of the peaks of absorption and pushed them to a smaller energy region since I has a bigger electronic influence than Br and Cl. The static polarization and band gap are linked inversely by the Penn's model[72]:

$$\varepsilon_1(0) \approx 1 + \left[\frac{\hbar\omega_p}{E_g}\right]^2$$

Where $\hbar$ and $\omega_p$ are the decreased plank constant and the frequency of the plasma, respectively.

The $\varepsilon_2(\omega)$ is significant because it determines the maximum absorption area and governs interband transitions that take place inside the materials used to fabricate devices. Due to the limitations of DFT, the transition points (VB to CB) exhibit a slight deviation from band structure and are plotted in Fig.6(b). The maximum value of $\varepsilon_2$ is called generally the first absorption peak (FAP) and is responsible for the electronic transition at the Fermi level. In the case of $A_2AuScCl_6$ (A= Cs/Rb), the FAP is 2.71 and 2.79 eV, respectively. The remaining compounds have various peaks in the range of 4 eV, which were seen for $A_2AuScBr_6$ (A= Cs, Rb) at 2.54 (2.56) and 2.83 (2.88) eV and for $A_2AuScI_6$ (A= Cs/Rb) at 2.27 (2.30), 2.79 (2.88), and 3.29 (3.35) eV, respectively. For the titled compounds, we simply need to know the highest FAP values, which are 2.71 (2.79) eV, 2.54 (2.56 eV), and 2.27 (2.30) eV for $A_2AuScX_6$ [A = Cs, Rb; X = Cl, Br, I], respectively. These values are in the visible range and indicate an effective absorption capacity. Compared to the DOS, the FAS originates mainly by the Au-$d$ along with X-$p$ at MVB (Maximum of VB) and the Sc-$d$ at MCB (Minimum of CB) point.

The absorption capacity $\alpha(\omega)$, is comparable to that of $\varepsilon_2(\omega)$ also illustrated by their coefficients at various energies (photon energy) and wavelength, as shown in Fig. 6(c) and Fig. 6(d). The formula for the absorption coefficient α(ω) was derived from the imaginary and real parts of the dielectric function[68].

$$\alpha(\omega) = 2\omega \left( \frac{[\varepsilon_1^2(\omega) + \varepsilon_2^2(\omega)]^{1/2} - \varepsilon_1(\omega)}{2} \right)^{1/2}$$

The absorption edge, also known as the optical band gap or resultant value $\alpha(\omega)$ of the material precisely corresponds to the electronic band gap, and the first highest absorption is observed at 2.95 (3.07), 2.67 (2.72), and 2.34 (2.36) eV for $A_2AuScX_6$ [A = Cs, Rb; X = Cl, Br, I], respectively, which are in the visible range. Therefore, these DPs function well as absorbents for the visible spectrum. Other maximum absorption peaks, which also lie in the visible to UV range, are found at 2.67 (2.72) and 3.03 (3.07) eV for $A_2AuScBr_6$ (A= Cs/Rb) and 2.34 (2.36), 3.11 (3.16) eV, and 3.36 (3.42) eV for $A_2AuScI_6$ (A= Cs/Rb). The development of several peaks with various peak intensities in the upper range from 7 to 13 eV illustrates the various possibilities of transition from filled to unfilled states.

Besides, these materials have superior absorption coefficients than others, such as 3 to $6.5\times10^4$ cm$^{-1}$ for $Cs_2CuBiX_6$ (X= Cl/Br/I)[73], 4.8 to $5\times10^4$ cm$^{-1}$ for $Rb_2CuBiX_6$ (X= Cl/Br)[74], 5 to $5.3\times10^4$ cm$^{-1}$ for $K_2CuBiX_6$ (X= Cl/Br)[75] and $8\times10^4$ cm$^{-1}$ for $Cs_2AgBiI_6$[76], according to the highest peaks of both compounds, which are situated at wavelength of visible range (400-700nm). The visible range absorption coefficients for $A_2AuScCl_6$ (A= Cs/Rb) is 3.33 (3.45) $\times$ 10$^5$ cm$^{-1}$. Other compounds show the different highest peak in the visible to UV range, for instance,

$A_2AuScBr_6$ (A= Cs/Rb) is 2.20 (2.24)× $10^5$ cm$^{-1}$ and 2.70 (2.81)× $10^5$ cm$^{-1}$, and $A_2AuScI_6$ (A= Cs/Rb) is 2.13 (2.18)×$10^5$ cm$^{-1}$, 3.17 (3.20) ×$10^5$ cm$^{-1}$ and 4.46 (4.55) ×$10^5$ cm$^{-1}$, respectively. Due to their excellent photoelectric defining features, these DP halide materials might be employed as the layer that absorbs sunlight in solar cells.

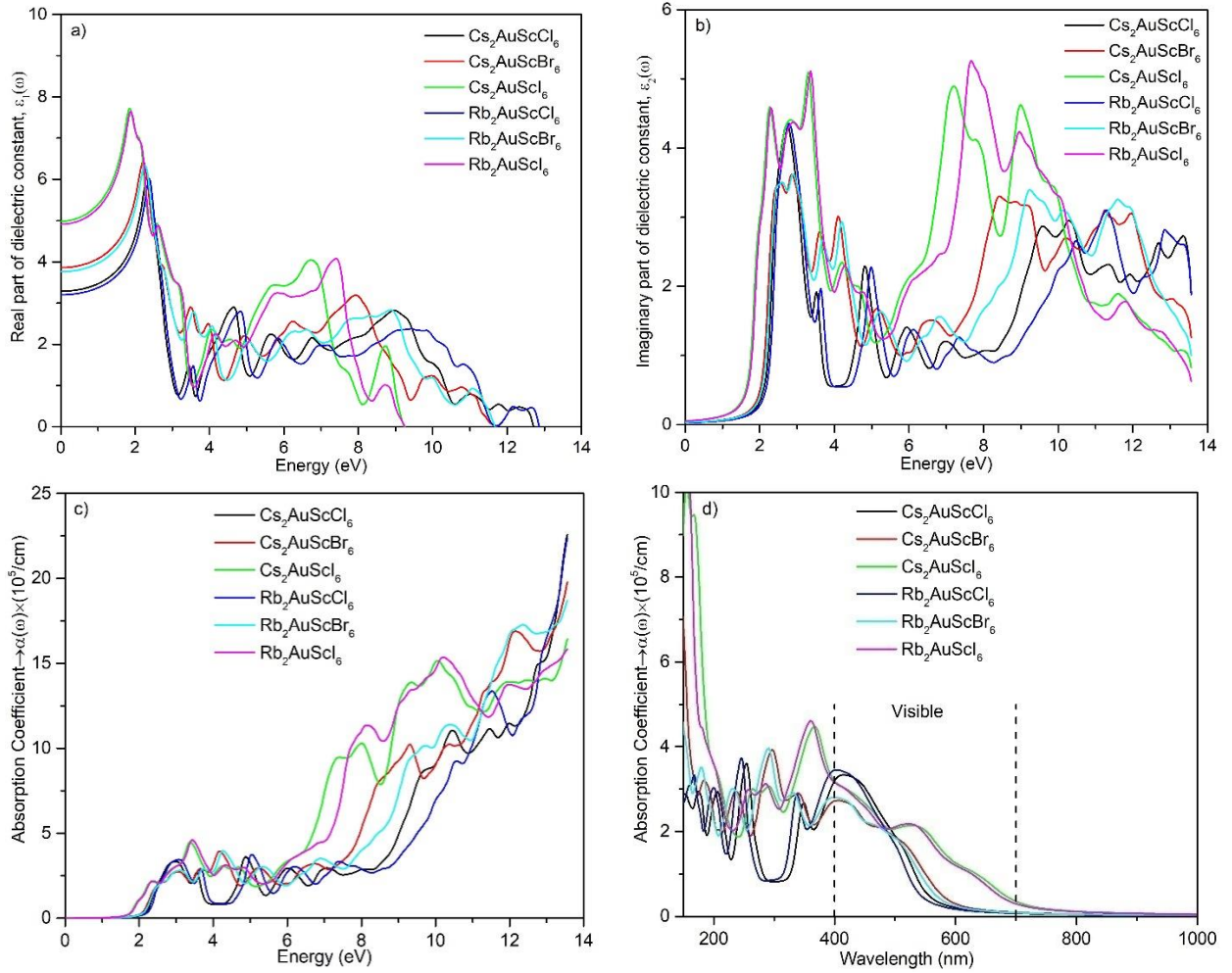

Fig.6: Obtained values of a) real part of the dielectric constant, b) imaginary part of the dielectric constant, c) absorption coefficient in eV and d) absorption coefficient in nm for DP of $Cs_2AuScCl_6$, $Cs_2AuScBr_6$, $Cs_2AuScI_6$, $Rb_2AuScCl_6$, $Rb_2AuScBr_6$, and $Rb_2AuScI_6$.

Other essential optical constants, such as the refractive factor n(ω), the extinction factor k(ω), the reflectivity R(ω), and loss factor L(ω), are all calculated from the dielectric properties of all compounds[77,78,79].

$$n(\omega) = \left[ \frac{\{\varepsilon_1^2(\omega) + \varepsilon_2^2(\omega)\}^{1/2}}{2} + \frac{\varepsilon_1(\omega)}{2} \right]^{1/2}$$

$$k(\omega) = \left[\frac{\{\varepsilon_1^2(\omega) + \varepsilon_2^2(\omega)\}^{1/2}}{2} - \frac{\varepsilon_1(\omega)}{2}\right]^{1/2}$$

$$R(\omega) = \frac{\{n(\omega) - 1\}^2 + k^2(\omega)}{\{n(\omega) + 1\}^2 + k^2(\omega)}$$

$$L(\omega) = \frac{\varepsilon_2}{\varepsilon_1^2 + \varepsilon_2^2}$$

The refractive index calculation results are displayed in Fig.7(a). The link between $n^2(\omega) = \varepsilon_1(\omega)$ caused the real part of the dielectric function and refractive index to behave in the same way[80]. The static refractive factors of n(0) are 1.81 (1.79), 1.96 (1.93) and 2.22 (2.20), and their maximum values are 2.49 (2.48), 2.57 (2.53) and 2.80 (2.78), with 2.28 (2.37) eV, 2.20 (2.28) eV, and 1.84 (1.87) eV, for $A_2AuScX_6$; A= Cs/Rb, X= Cl/Br/I), respectively. Optoelectronic activities are appropriate for n(ω) values from 2.0 to 4.0[81], and the studied materials have values within this range. The fact that the observed refractive factor values are more than 1 is strong evidence for semiconducting materials. The refractive factor also evaluates the strength of the bonds—whether they are more or less than unity. A refractive index greater than one indicates the presence of covalent bonds rather than ionic ones. Due to photon slowing and increasing electron density, the aforementioned materials displayed covalent bonding, either static or dynamic[37].

Table 4: Computed optical constants.

| Compound | $\varepsilon_1(0)$ | n(0) | R(0) |
|---|---|---|---|
| $Cs_2AuScCl_6$ | 3.27 | 2.22 | 8.37% |
| $Cs_2AuScBr_6$ | 3.85 | 1.96 | 10.70% |
| $Cs_2AuScI_6$ | 4.98 | 1.81 | 14.58% |
| $Rb_2AuScCl_6$ | 3.19 | 2.20 | 7.96% |
| $Rb_2AuScBr_6$ | 3.74 | 1.93 | 10.19% |
| $Rb_2AuScI_6$ | 4.90 | 1.79 | 14.07% |

The identical pattern of $\varepsilon_2(\omega)$ and $\alpha(\omega)$ is followed by the extinction factor k(ω), which detects the loss of EM waves in contents, as shown in Fig.7(b). The connection between k(w) and α(w) is as follows:

$$k = \frac{\alpha\lambda}{4\pi}$$

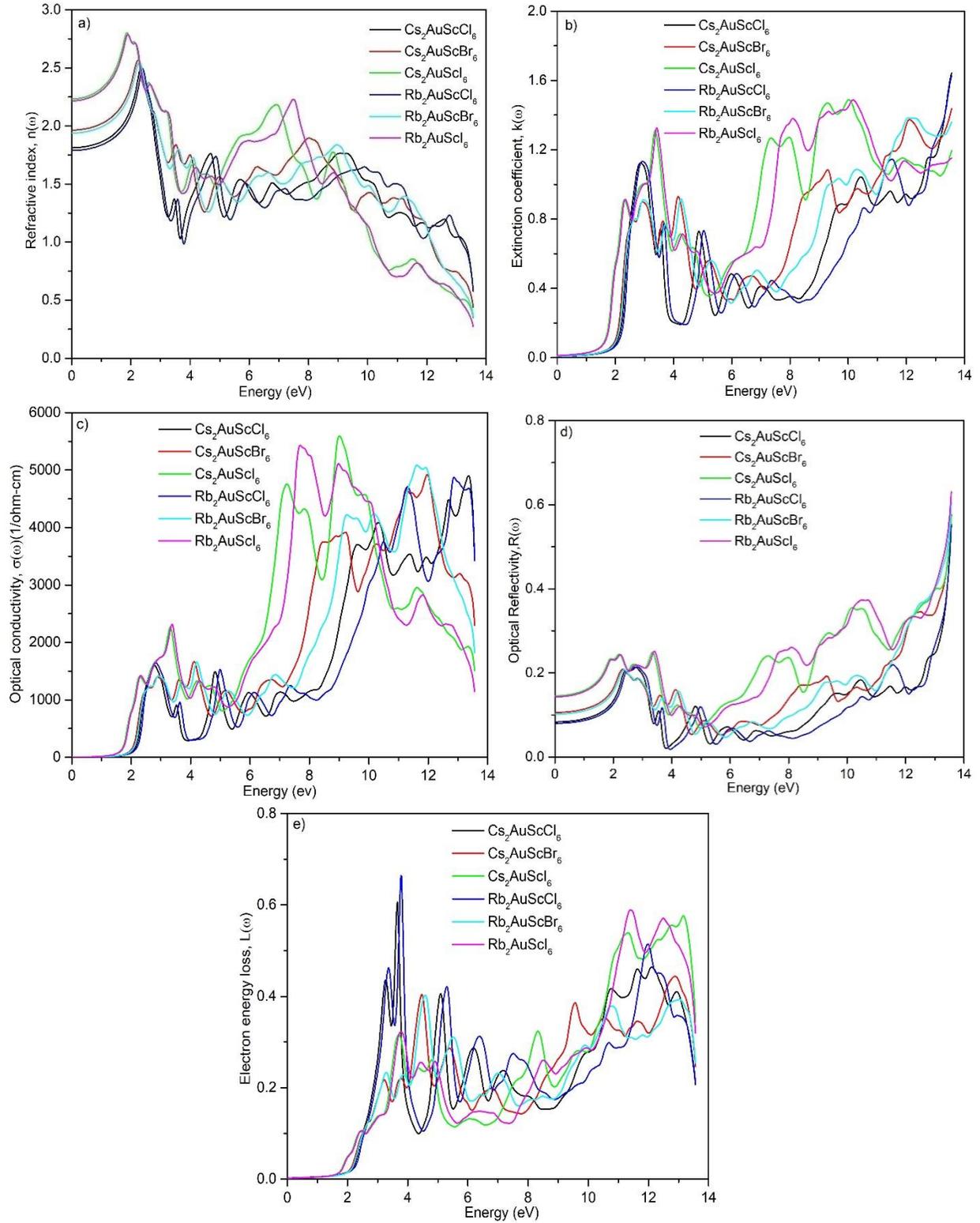

Fig.7: The obtained a) refractive index, b) extinction coefficient, c) optical conductivity d) optical reflectivity, and e) electron energy loss of $Cs_2AuScCl_6$, $Cs_2AuScBr_6$, $Cs_2AuScI_6$, $Rb_2AuScCl_6$, $Rb_2AuScBr_6$, and $Rb_2AuScI_6$.

The value of the extinction coefficient was discovered to be extremely low, as would be predicted for semiconductor materials[82]. Any material's surface will absorb some of the incident light if a stream of photons at a specific frequency strikes it, initiating the interband transition or switching of electrons. The amount of real optical conductivity is used to measure this change. For optoelectronic devices to function practically, the optical conductivity, which is determined by the interband movement of electrons, must be within 1.4 to 4.0 eV in the visible range[81]. The initial peak exhibits maximum conductivities of 1594/1637 $(\Omega\text{-cm})^{-1}$ at 2.71/2.81 eV for $A_2AuScCl_6$ (A= Cs/Rb), 1227/1247 $(\Omega\text{-cm})^{-1}$ at 2.58/2.64 eV and 1401/1412 $(\Omega\text{-cm})^{-1}$ at 2.88/2.90 eV for $A_2AuScBr_6$ (A= Cs/Rb), and 1402/1417 $(\Omega\text{-cm})^{-1}$ at 2.29/2.31 eV, 1749/1755 $(\Omega\text{-cm})^{-1}$ at 2.99/3.01 eV and 2266/2299 $(\Omega\text{cm})^{-1}$ at 3.29/3.35 eV for $A_2AuScI_6$ (A= Cs/Rb), respectively, while the highest value observed (4100 to 5500 $1/\Omega\text{-cm}$) from 7 to 13 eV, as shown in Fig. 7(c).

The static value of reflectivity is less than 15% for the explored compounds. The lowest reflectivity $R(\omega)$ indicates that both materials are thought to be more effective for incoming light, with energy values ranging from zero to the gap in the band, as shown in Fig.7(d). The computed value of $L(\omega)$ measures the amount of energy (less than 0.3 in the visible range) that disappears when a photon travels through a substance, as shown in Fig.7(e). A very low reflectivity and loss function value is significant for the materials to be used as absorbing layers in solar cells. Finally, these materials can be used in optoelectronic gadgets because of their adjustable computing band gap, strong dielectric strength, excellent absorbance spectra, and good photoconductivity.

### 3.5. Thermoelectric properties

Given the increasing demand for energy, waste heat recovery—a method that transforms heat into electrical energy—becomes increasingly important in systems to improve material efficiency. Thermoelectric (TE) materials produce portable power for distant missions, energy devices, and cooling applications. An analysis of a compound's transport properties is necessary to determine whether it is appropriate for these uses. For the materials under study, effective thermoelectric conversion requires reducing heat conductivity ($\kappa$) and increasing electrical conductivity ($\sigma$). The See-beck coefficient (S), power factor (PF), and figure of merit (ZT) would all rise as a result of this. Additional information, such as a smaller band gap and greater carrier mobility must be considered when analyzing TE materials.

These materials' electronic band structure and phonons' dispersion—vibrations that move heat across the lattice—determine their thermoelectric (TE) properties. Lattice thermal vibrations ($K_L$) are calculated in the thermo-mechanical section (supplementary data, Table S1).

To calculate the TE behavior with respect to temperature, we employ the Boltzmann transport theory[83]. Specifically, we utilize the Boltztrap2 code[45] for integrating Fermi energy, a process dependent on the concentration of both n and p-type charge carriers. Additionally, we incorporate the electronic components of thermal conductivity (Ke) and electrical conductivity (σ) into the calculation, considering a relaxation time ($\tau = 10^{-14}$s).

Table 5: Room temperature computed values of electrical conductivity (σ), electronic conductivity ($K_e$), See-beck coefficient (S), power factor (PF), and figure of merit (ZT) of the investigated compounds.

| Compound | σ (×10$^5$ /Ω.m) | Ke (×10$^{14}$ Wm$^{-1}$K$^{-1}$) | S (mV/K) | PF (×10$^{-3}$ Wm$^{-1}$K$^{-2}$) | ZT |
|---|---|---|---|---|---|
| Cs$_2$AuScCl$_6$ | 1.06 | 0.43 | 0.192 | 3.98 | 0.92 |
| Cs$_2$AuScBr$_6$ | 0.90 | 0.37 | 0.211 | 4.03 | 1.07 |
| Cs$_2$AuScI$_6$ | 0.87 | 0.37 | 0.213 | 4.01 | 1.06 |
| Rb$_2$AuScCl$_6$ | 1.01 | 0.41 | 0.199 | 4.02 | 0.97 |
| Rb$_2$AuScBr$_6$ | 0.90 | 0.39 | 0.208 | 3.92 | 0.99 |
| Rb$_2$AuScI$_6$ | 0.92 | 0.39 | 0.209 | 4.06 | 1.01 |

Electrical conductivity (σ) measures how well a material conducts electricity. The initial electrical conductivity values at 200K for Cs$_2$AuScX$_6$ (Rb$_2$AuScX$_6$), where X represents Cl, Br, or I, are 0.73 (0.66) × 10$^5$, 0.56 (0.55) × 10$^5$, and 0.54 (0.57) × 10$^5$ (1/Ω.m), respectively. A key feature of semiconductors is that all compounds increase electrical conductivity with temperature[84]. Moreover, the I-based compound outperforms the Br and Cl-based compounds, achieving the greatest conductivity values as the temperature becomes closer to its maximum range of 200 to 1000K, as shown in Fig. 8(a). This is explained by the fact that more bonds are broken in halides based on Cs and Rb. The electrical conductivity values that were measured at

high temperatures (1000K) are 2.27 (2.29), 2.38 (2.39), and 2.53 (2.58) × $10^5$ (1/Ω.m) for $Cs_2AuScCl_6$ ($Rb_2AuScCl_6$), $Cs_2AuScBr_6$ ($Rb_2AuScBr_6$), and $Cs_2AuScI_6$ ($Rb_2AuScI_6$), respectively.

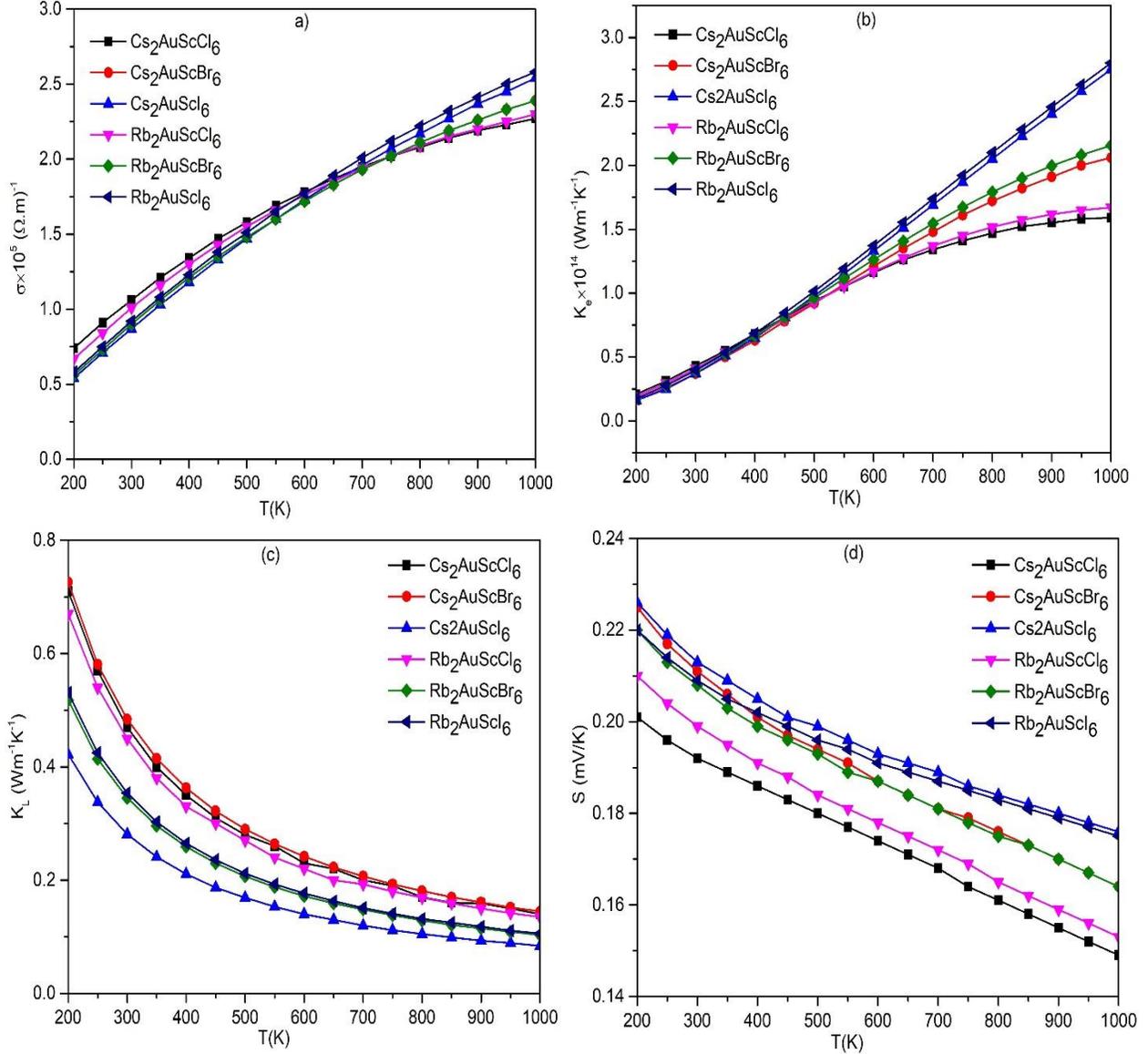

Fig.8: The computed a) Electrical conductivity, σ, b) Electronic part of thermal conductivity, $K_e$, c): Lattice thermal conductivity (from supplementary part) $K_L$, and d) See-beck coefficient, $S$, of the studied compounds.

The trends in the graphs of $K_e$ and σ versus temperature (T) are roughly the same. The electronic conductivity increases with temperature. As can be seen in Fig. 8(b), the highest values of 1.58 (1.66), 2.05 (2.15), and 2.75 (2.79) $Wm^{-1}K^{-1}$ are obtained at 1000K for $Cs_2AuScCl_6$

($Rb_2AuScCl_6$), $Cs_2AuScBr_6$ ($Rb_2AuScBr_6$), $Cs_2AuScI_6$ ($Rb_2AuScI_6$), respectively. The low thermal conductivity ($\kappa$), which is the sum of $K_L$ and $K_e$[85,86], and high σ values of these materials make them appropriate for thermoelectric applications. Table 5 provides the computed parameters for σ, $K_e$, and $K_L$ at room temperature (300K). Interestingly, their electrical conductivity (σ) is a factor of $10^5$ higher than their heat conductivity ($K_e$), highlighting their considerable potential for optoelectronic applications.

When applying a temperature gradient, the See-beck coefficient (S) measures the voltage produced across a material. In thermoelectric materials, a high See-beck coefficient is highly desirable[87]. Notably, the See-beck coefficient decreases as the temperature rises. In the case of Cs (Rb)-based compositions containing Cl, Br, and I atoms, the S values decrease from 0.201 to 0.149 (0.231 to 0.152) mV/K, 0.224 to 0.163 (0.249 to 0.164) mV/K, and 0.226 to 0.175 (0.243 to 0.174) mV/K, respectively, over the temperature range of 200 to 1000K, as displayed in Fig.8(d). The S values observed at 300K are given in Table 5 and exhibit the following trend: $Cs_2AuScI_6$ > $Cs_2AuScBr_6$ > $Cs_2AuScCl_6$ for Cs-based halides and $Rb_2AuScI_6$ > $Rb_2AuScBr_6$ > $Rb_2AuScCl_6$ for Rb-based halides. Because these compounds have positive S values within the designated temperature range are classified as P-type.

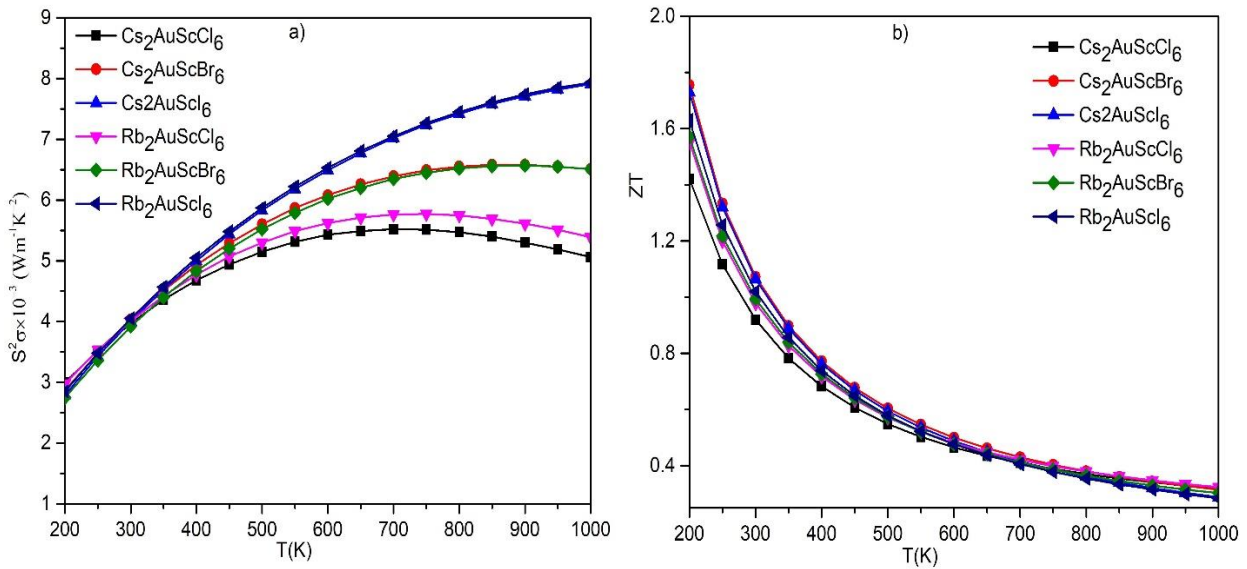

Fig.9: a) Power factor (PF) $S^2\sigma$ and b) Figure of merit (ZT) of the studied compounds.

One of the most important factors in assessing a material's TE characteristics is its power factor. It measures how well a thermoelectric material transforms heat into electrical power. In mathematics, the power factor is defined as[88]: PF = $S^2\sigma$, where $S$ represents the See-beck coefficient, also known as thermo-power or thermoelectric voltage, and $\sigma$ denotes the electrical conductivity. The graph illustrating PF versus temperature (T) can be seen in the Fig.9 (a). As temperature increases, the power factor also rises, reaching its peak values of 5.51 Wm$^{-1}$K$^{-2}$, 6.55 Wm$^{-1}$K$^{-2}$, and 7.88 Wm$^{-1}$K$^{-2}$ at 700K, 800K, and 1000K for $Cs_2AuScCl_6$, $Cs_2AuScBr_6$, and $Cs_2AuScI_6$, respectively. For $Rb_2AuScCl_6$, $Rb_2AuScBr_6$, and $Rb_2AuScI_6$, the corresponding values are 5.76 Wm$^{-1}$K$^{-2}$, 6.54 Wm$^{-1}$K$^{-2}$, and 7.90 Wm$^{-1}$K$^{-2}$ at 700K, 800K, and 1000K, respectively. However, the studied compounds have higher PF at room temperature than some known TE materials, such as 2.64 and 2.61 for $Tl_2(Se, Te)Cl_6$[89].

The *ZT* value is a dimensionless figure of merit that assesses the effectiveness of the TE materials in association with the aforementioned characteristics. Measurement of the dimensionless figure of merit is used to measure thermoelectric material efficiency. *ZT* is defined as $ZT = \frac{s^2\sigma}{\kappa}T$, where κ is the thermal conductivity made up of the electronic and lattice parts, $\sigma$ is the electrical conductivity, and *S* is the thermo-power. Promising TE materials are those with a high ZT~1.0, which may be attained when the power factor $s^2\sigma$ is high, and the thermal conductivity is low[90]. Nevertheless, it is important to remember that a rise in temperature causes the *ZT* value to decrease. As seen in Fig. 9(b), the *ZT* values at 300K for compositions based on Cs and including Cl, Br, and I atoms are 0.92, 1.07, and 1.06, respectively, whereas for compositions based on Rb and containing Cl, Br, and I atoms, they are 0.97, 0.99, and 1.01, respectively. So these materials show higher *ZT* values compared with other DP halide such as 0.77, 0.82, and 0.95 for $Rb_2ScTlX_6$ (X = Cl, Br, and I)[81], 0.63, 0.61, and 0.66 for $Cs_2ScTlX_6$ (X = Cl, Br, I)[38], 0.61, 0.72, and 0.81 for $Cs_2KTlX_6$ (X= Cl, Br, I)[91], 0.76 to 0.78 for $Rb_2AlInX_6$ (X= Cl, Br, I)[92] and the most studied DP halide is about 0.86 and 0.84, for $Cs_2AgBiX_6$ (X= Cl, Br)[93]. Also, the studied two halides, $Cs_2AuScBr_6$ and $Cs_2AuScI_6$, can be considered more promising TE materials (*ZT* = 1.07 and 1.06) than $Cs_2PtI_6$ (*ZT* = 1.03)[85] due to high See-beck coefficient values at room temperature. Therefore, the studied DP materials could fulfill the necessity for waste heat use and sustainable energy solutions.

## 4. Conclusion

The ground state structure is obtained for the titled compounds by the DFT calculations with the aid of the Wien2k code. The structural stability is confirmed by the formation energy, binding energy, phonon dispersion curve, stiffness constants, tolerance, and octahedral factors. The lattice constants increase from 10.44 to 11.89Å when Br and I replace the Cl ion. The electronic band structure confirms the semiconducting nature. Using the TB mBJ potential, the energy band gaps of $A_2AuScCl_6$ (A= Cs/Rb), $A_2AuScBr_6$ (A= Cs/Rb), and $A_2AuScI_6$ (A= Cs/Rb) are 1.88 (1.93), 1.68 (1.71), and 1.30 (1.32) eV, respectively. The band gap values of the DP of the titled compounds are close to the S. Q. limit[94] for maximum efficiency, which makes them suitable for application in solar cell devices. The lowest effective mass of an electron and a hole that satisfies green or sustainable energy technology. The DOS and charge density mapping plots explore the bonding nature present within the studied materials. The important optical constants demonstrate the studied material's response to the incident photon. The usefulness of the titled compounds in optoelectronic applications stems from their high absorption coefficients of $10^5$ order and other optical constants, including low energy loss and reflection capacity (reflectivy is less than 15%). The DPs, $A_2AuScX_6$ (A= Cs, Rb; X= Cl, Br, I), are, therefore, promising options for efficient solar cells and reasonably priced optoelectronic devices because of their strong light absorption capacity and remarkable adjustable band gap. Analysis of thermoelectric properties reveals essential prospects for the titled compounds. The reduced lattice thermal conductivity and band gap values confirm their usage in thermoelectric devices. The high ZT values of 0.92, 1.07, 1.06, 0.97, 0.99, and 1.01 calculated power factor and thermal conductivity at 300K for $Cs_2AuSccl_6$, $Cs_2AuBr_6$, $Cs_2AuScI_6$, $Rb_2AuScCl_6$, $Rb_2AuScBr_6$, and $Rb_2AuScI_6$, respectively imply that studied compounds are promising thermoelectric materials. The elastic properties certify the ductility that confirms the studied compound's machinability, which is essential for device fabrication. Moreover, high melting points and low $K_{min}$ values also support their high-temperature use. In the end, the obtained results of the titled DPs are encouraging and hope to provide proper guidance for future research on energy harvesting materials.

## Declaration of interests

The authors declare that they have no known competing financial interests or personal relationships that could have appeared to influence the work reported in this paper.

**Acknowledgments**

This work was carried out with the aid of a grant (grant number: 21-378 RG/PHYS/AS_G - FR3240319526) from UNESCO-TWAS and the Swedish International Development Cooperation Agency (SIDA). The views expressed herein do not necessarily represent those of UNESCO-TWAS, SIDA or its Board of Governors.

**CRediT Author contributions**

**S. Mahmud:** Conceptualization, Formal analysis, Methodology, Data calculations, Validation, Writing – original draft. **M.A.Ali:** Conceptualization, Methodology, Formal analysis, Validation, Project administration, Writing – original draft, Supervision. **M.M. Hossain:** Writing – review & editing, Validation. **M.M. Uddin:** Writing – review & editing, Validation.